# Multifractality of Drop Breakup in Air-blast Nozzle Atomization Process

Wei-Xing Zhou[1] and Zun-Hong Yu

*East China University of Science and Technology, P.O. Box 272, Shanghai 200237, P.R.China*

The multifractal nature of drop breakup in air-blast nozzle atomization process has been studied. We apply the multiplier method to extract the negative and the positive parts of the $f(\alpha)$ curve with the data of drop size distribution measured using Dual PDA. A random multifractal model with the multiplier triangularly distributed is proposed to characterize the breakup of drops. The agreement of the left part of the multifractal spectra between the experimental result and the model is remarkable. The cause of the distinction of the right part of the $f(\alpha)$ curve is argued. The fact that negative dimensions arise in the current system means that the spatial distribution of the drops yielded by the high-speed jet fluctuates from sample to sample. On other words, the spatial concentration distribution of the disperse phase in the spray zone fluctuates momentarily showing intrinsic randomness.

**PACS** number(s): 47.27.Wg, 47.53.+n, 47.55.Kf, 02.50.Ey

## I. INTRODUCTION

The transformation of bulk liquid into sprays and other physical dispersions of small droplets in a gaseous atmosphere is of importance in many agricultural and industrial processes, which includes applying agricultural chemicals to crops, paint spraying, spray drying, food processing, cooling of nuclear cores, combustion, gasification, and some other fields. For sake of complexity of atomization process, it is too difficult to disclose clearly the mechanism and also impossible to combine in one model all the influence factors, such as equipment dimensions, size and geometry of nozzle, physical properties of the dispersed phase and the continuous phase, and operating mode. Hence, a lot of empirical and semi-empirical models have been established so far to describe atomization processes [1]. To elaborate on the distortion and breakup mechanisms of a single liquid droplet injected into a transverse high-velocity air jet, diverse regimes were presented [2-6], which ignores the interaction of multiple droplets.

According to Mandelbrot [7], there exist a scaling principle of drop size distribution in many physical systems. In Ref. [8], the fractal facet of drop size distribution was studied with two characteristic parameters, the textural and structural fractal dimensions [9]. However, it is not sufficient that only two fractal dimensions are investigated to characterize the singularities of size distribution. The distribution of singularities on the geometry support, which is described by the multifractal spectrum $f(\alpha)$, should be considered [10,11]. Moreover, since there exists the intrinsic or practically induced randomness in many real systems, the application of the theory of random multifractals is indispensable, such as turbulence [12] and DLA [13]. Negative dimensions appear in these cases [14,15], which is a brand of randomness in many cases[2].

As always, the measure considered in the theory of random multifractals is constructed through a randomly multiplicative cascade process. Let a cascade begin with mass equal to 1, uniformly spread over [0,1], and let the *n*th cascade stage share at random the mass in a cell of length $b^{-n}$ among $b$ sub-cells of length $b^{-n-1}$ with multipliers $M_i$ ($1 \leq i \leq b$). Since mass is conserved while it is spread around within a cell, strong conservation rules constrain the multipliers. Here the multipliers $M_i$ are identically distributed with p.d.f $\Pr(M)$ and independent. From Cramer's theorem of large deviations [16,17], the mass exponent, which is defined by an annealed averaging of moments of the multipliers, namely

$$\tau(q) = -D_0 - \frac{\log\langle M^q \rangle}{\log b}, \tag{1}$$

is linked with the multifractal spectrum $f(\alpha)$ by the Legendre and inverse Legendre transform [18]. Here, the fractal dimension $D_0$ is equal to 1. Therefore, we have

$$\alpha(q) = \tau'(q) = -\frac{\langle M^q \log M \rangle}{\langle M^q \rangle \log b}, \tag{2}$$

and

$$f(\alpha(q)) = q\alpha(q) - \tau(q) = \frac{\langle M^q \rangle \log\langle M^q \rangle - \langle M^q \log M^q \rangle}{\langle M^q \rangle \log b} + D_0. \tag{3}$$

---

[1] Corresponding author. Email address: wxzhou@ecust.edu.cn.

[2] In fact, negative dimensions may be absent in random multifractal measures.

In this paper, we will investigate the multifractal nature of the drop breakup in the process of air-blast nozzle atomization, which is the consecutive work of Ref. [8].

## II. MULTIFRACTAL NATURE OF DROP BREAKUP
### A. Experimental

The measurement of the drop size distribution is carried out using the 58N81 Dual Particle Dynamic Analyzer (Dual PDA), which is based on the Doppler effect. The focus of the lens used in the experiment is 500mm. The power of the argon ion laser generator is 5W, which generate three laser beams with colors green, blue and purple. The minimal increment of the movement of the measurement point is 0.02mm. The angle included between the two arms is 63 degree. The schematic chart of the measurement system is illustrated in Fig.1.

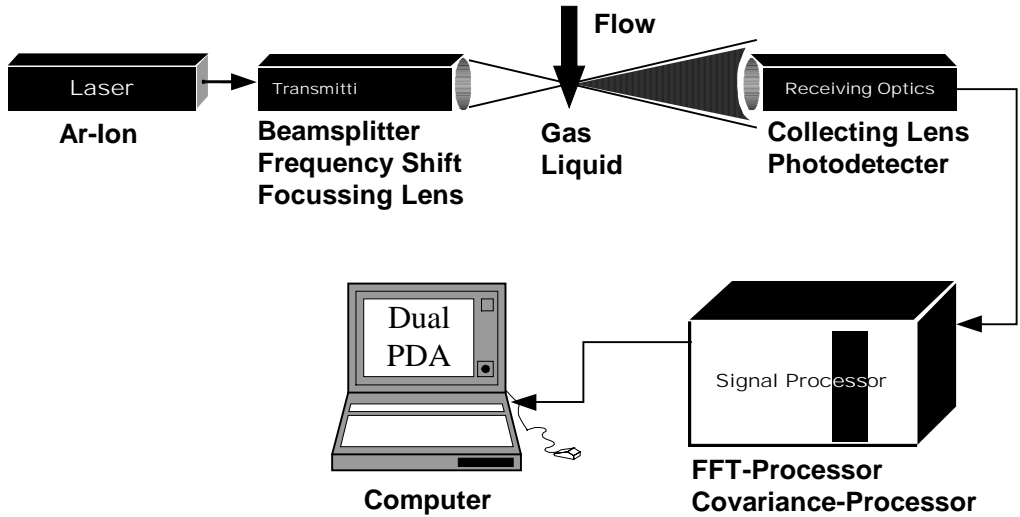

FIG. 1 Schematic chart of the Dual PDA system

The diameter of the central passage of the selected nozzle is 3mm, while the width of the annular space is 11.5mm. The rake angle of the nozzle is 10 degree. The liquid phase water passes through the central passage with the flow rate of $0.21\text{m}^3/\text{h}$, while the continuous phase air speeds through the annular space with $115\text{m}^3/\text{h}$ in flow rate. The measurement is carried out at different spatial positions throughout the spray zone, which covers 324 points with the distance of 10 mm up to 440 mm. The number of validated samples is up to 2000.

A record with 525287 data points was obtained using Dual PDA sketched in Fig. 1. A portion of the diameter signals of the droplets in the spray zone is illustrated in Fig. 2. The Hurst exponent of the total record is about $H = 0.81$, which show a relatively strong persistence. Therefore, the mass change of the adjacent drops is not large and one can expect $\Pr(M)$ to have a relatively great proportion near $M = 0.5$.

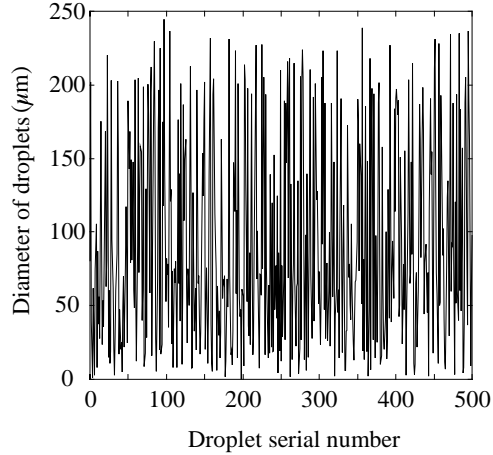

FIG. 2. Portion of the diameter signals of the droplets in the spray zone. Note that the signals are unilateral and the Hurst exponent is about 0.81. Hence, the mass change of the adjacent drops is not large and one can expect that $\Pr(M)$ have a relatively great proportion near $M = 0.5$.

## B. Multifractal nature obtained by using multiplier method

In order to extract the spectrum of the singularities $f(\alpha)$ from the experimental data, one can use the procedure from the scaling of histograms of multifractal measures, which is a box-counting method in essence [19]. This method works well when the scaling range is wide, and useful statistical information on iso-$\alpha$ set such as lacunarity can be obtained. However, if the scaling range is small, corrections of high order become necessary, which leads to exponentially more work. In order to improve the accuracy and reduce the work when computing the $f(\alpha)$ function, the multiplier method should be utilized [20-23]. This approach is based on the viewpoint that the scaling properties reflecting the self-similar structure of the measure can be described in terms of a repeated composition of a level-independent distribution of multipliers that define the rearrangement of the measure into small pieces [12,18]. By suitably manipulating these distributions one can compute the complete $f(\alpha)$ function from Eq. (3). Hence, to compute the $D_q$, $\tau(q)$ or $f(\alpha)$ function is to obtain the multiplier density $\Pr(M)$.

The random multifractal measure measured in the experiment is composed of 525287 pieces with equal scale on the interval $(0,1)$. The mass distributed on the $i$ th piece equals to the mass of the corresponding drop $m_i$. The simplest way of computing $\Pr(M)$ is to cover the measure at the $n$ th generation with boxes of size $l = 2^{1-n}$ and evaluate the mass in each box, then subdivide each of these boxes into two pieces and compute the ratios of the masses $M$ in the original box to any one of the two subdivided pieces. A relatively accurate way is to evaluate the averaging in Eqs. (1-3) in a discrete form. Thus we have

$$\langle g(M) \rangle = 2^{1-n} \cdot \sum_{i=1}^{2^{n-1}} [g(M_i) + g(1-M_i)]. \tag{4}$$

The dotted lines shown in Figs.3-6 are the mass exponents $\tau(q)$, generalized dimensions $D_q$, singularity strength $\alpha(q)$ and multifractal spectrum $f(\alpha)$, respectively.

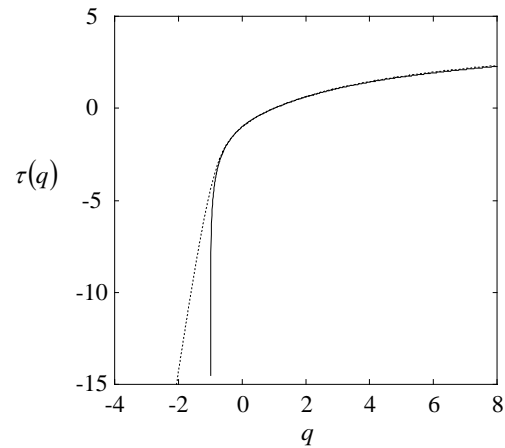

FIG. 3. Comparison between the mass exponents obtained from the multiplier method (dotted lines) and from the randomly multiplicative cascade model (solid lines) for drop breakup in the atomization process.

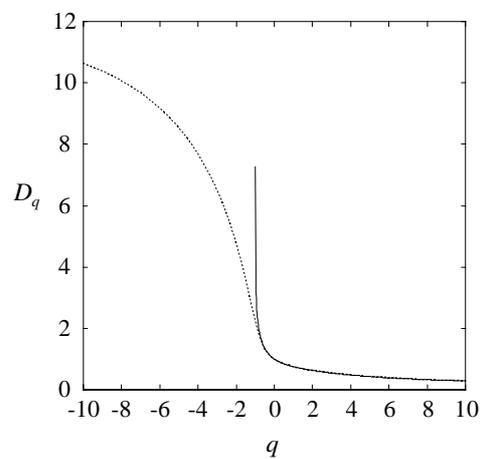

FIG. 4. Comparison between the generalized dimensions obtained from the multiplier method (dotted lines) and from the randomly multiplicative cascade model (solid lines) for drop breakup in the atomization process.

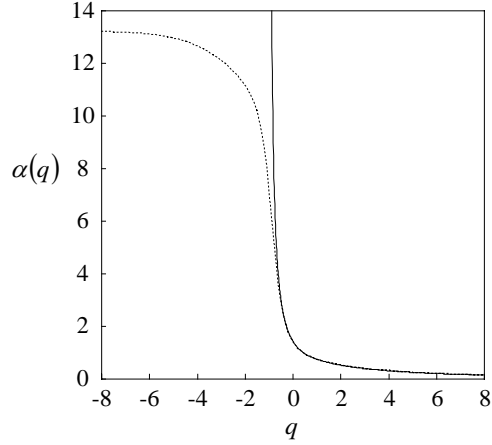

FIG. 5. Comparison between the strengths of singularities obtained from the multiplier method (dotted lines) and from the randomly multiplicative cascade model (solid lines) for drop breakup in the atomization process.

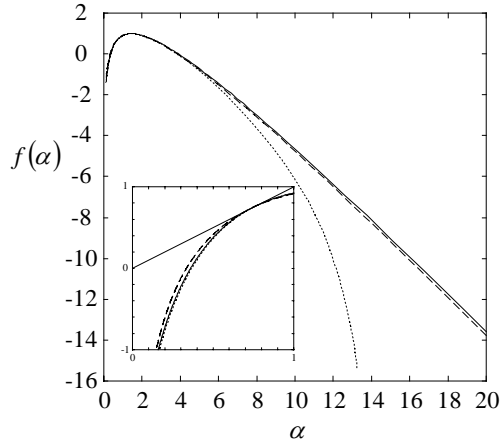

FIG. 6. Comparison between the multifractal spectra obtained from the multiplier method (dotted lines) and from the randomly multiplicative cascade model (solid lines) for drop breakup in the atomization process. The two results agree remarkably with each other for $q \geq 0$. However, when $q$ approaches $-1^+$, the discrepancy between the two methods becomes more and more notable. The dashed line is from the random model with $\Pr(M) = 1$.

## C. Randomly multiplicative cascade model for drop breakup in atomization process

Let $N(\mathrm{d}M)$ denotes the number of multipliers lying in the interval $\left(M - \frac{1}{2}\mathrm{d}M, M + \frac{1}{2}\mathrm{d}M\right]$. The probability density of the multipliers can be calculated by

$$\Pr(M) = \frac{N(\mathrm{d}M)}{\sum N(\mathrm{d}M)} \bigg/ \mathrm{d}M . \tag{5}$$

Note that $\sum N(\mathrm{d}M) = 2^n$. We plot all $\Pr(M)$ at different spatial position in the spray zone with $\mathrm{d}M = 0.001$ in Fig. 7. One can find that these p.d.f's of multipliers at different position agree with each other and are hence stable in the spray zone, which imply that drop breakup is homogenous throughout the spray zone in the viewpoint of probability distribution of multipliers. Since the multipliers from the same mother box are $M$ and $1-M$, the resulting multiplier distribution must be symmetrical, that is

$$\Pr(M) = \Pr(1-M). \tag{6}$$

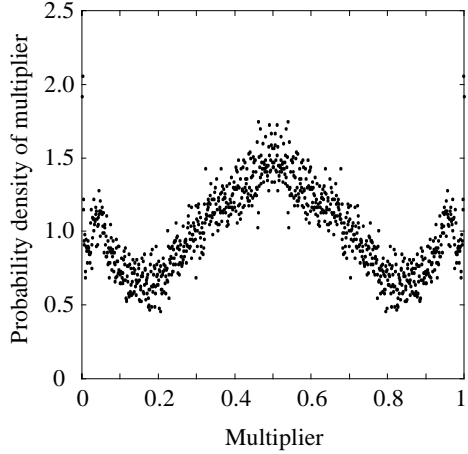

FIG. 7. The probability density $\Pr(M)$ of the multiplier measured from the experiment of atomization. The multiplier distribution is symmetrical to the line $M = 0.5$. The unit interval $(0,1)$ is divided uniformly in 1000 sub-intervals. One can fit $\Pr(M)$ roughly with a piecewise triangular function, which leads to the analytic results of scaling properties.

We find that the empirical probability density $\Pr(M)$ is roughly consistent with the triangular form

$$\Pr(M) = \begin{cases} -3.6719M + 1.1875, & 0 < M \leq 0.16 \\ 2.6471M + 0.1765, & 0.16 < M \leq 0.5 \\ -2.6471M + 2.8235, & 0.5 < M \leq 0.84 \\ 3.6719M - 2.4844, & 0.84 < M < 1 \end{cases} \tag{7}$$

Hence, the averaged $q$-order moments are in the form

$$\langle M^q \rangle = \frac{1}{q+2}\left[-6.3189 \times 0.16^{q+2} + 5.2941 \times 0.5^{q+2} - 6.3189 \times 0.84^{q+2} + 3.6719\right]$$

$$+ \frac{1}{q+1}\left[1.0110 \times 0.16^{q+1} - 2.6471 \times 0.5^{q+1} + 5.3079 \times 0.84^{q+1} - 2.4844\right]. \tag{8}$$

It is obvious that $q_{\text{bottom}} = -1$. The reason is presented in section III. Similarly, we obtain the functions characterizing scaling properties that are shown in Figs.3-6 as solid lines. The two results agree remarkably with each other for $q \geq 0$. However, when $q$ approaches $-1^+$, the discrepancy between the two methods becomes more and more notable. The dashed line in Fig.6 is from the random multifractal model with $\Pr(M) = 1$.

### III. DISCUSSION

#### A. The lower limit of $q$ in the model

In the randomly multiplicative cascade model presented in section II(c), we find that there exists a lower limit $q_{\text{bottom}} = -1$ of parameter $q$. As a matter of fact, for most of random multifractal process, $q_{\text{bottom}}$ exists and is a finite value that may be either negative or zero, which depends upon the probability distribution of multipliers. For instance, consider a multiplicatively generated random measure whose probability density of multipliers satisfies a power law, namely $\Pr(M) = (x+1)M^x$, where $x > -1$. The lower limit is a negative number $-1-x$, and when $x$ tends to $-1^+$, $q_{bottom}$ approaches $0^-$. Recall that $q_{bottom}$ is also zero in left-sided multifractals [24,25]. However, these two cases are essentially different.

Assuming that $\Pr(M)$ can be expanded into Taylor series at $M = 0$, we have

$$\Pr(M) = \Pr(0) + \Pr'(0)M + \frac{1}{2!}\Pr''(0)M^2 + \frac{1}{3!}\Pr'''(0)M^3 + \cdots \tag{9}$$

Suppose that $\Pr^{(i)}(0) = 0$ for $i < n$ and $\Pr^{(n)}(0) \neq 0$. Then, the averaged moment can be expressed in the form

$$\langle M^q \rangle = \sum_{i=n}^{\infty} \frac{\Pr^{(i)}(0)}{i!} \int_0^1 M^{q+i} \, dM . \tag{10}$$

Since $\int_0^1 M^{q+i} \, dM$ has a spot at $M = 0$ for $q \leq -(i+1)$, one can evaluate the spot integral as the limit of the corresponding normal integral, namely

$$\int_0^1 M^{q+i} \, dM = \lim_{\delta \to 0^+} \int_\delta^1 M^{q+i} \, dM$$

$$= \begin{cases} \lim_{\delta \to 0^+} \frac{1 - \delta^{q+i+1}}{q+i+1}, & q < -(x+1) \\ \lim_{\delta \to 0^+} -\log \delta, & q = -(x+1) \end{cases}. \tag{11}$$

We find that $\langle M^q \rangle$ is undefined for $q \leq -(i+1)$, since the limit in Eq. (11) tends to $\infty$. By a transformation of $M = e^x$, it is easy to show that $\langle M^q \log M \rangle$ is only defined for $q > -(i+1)$ as well.

Now consider the case represented in Eq. (7). The integral concerning $\langle M^q \rangle$ has a spot at $M = 0$ and it is obvious that $i = 0$. Hence $q_{bottom}$ is identical to $-1$.

## B. The case of $q \geq 0$ and comparison with uniform distribution

We can find that, in Figs.3-6, the agreement between the experiment and the model is perfectly good for $q \geq 0$. We also presented an alternative model with $\Pr(M) = 1$ as shown in Fig.6 The cause is from the breakup mechanism of drops. We have estimated the Weber numbers of drops throughout the spray region according to the experiments. A majority of drops in the spray have the Weber numbers less that the critical value of $12$, indicating that these drops lie in the vibrational breakup regime [2]. Meanwhile, the rest drops have the Weber number between $12$ and $21.5$ falling in the bag breakup regime. In the vibrational breakup regime, one drop splits into two droplets with the mass ratio of droplet to its mother drop around $0.5$, while in the bag breakup regime, one drop splits into several relatively bigger droplets and many smaller droplets. Therefore, vibrational breakup dominates and bag breakup also arises. In the case of bag breakup, we can regard the mother drop as several dummy drops. These breakup regimes conform to the probability distribution of multipliers. Since a triangular distribution is more precise than the uniform distribution, the multifractal spectrum of the former coincides with that of the experimental result better that the latter.

Consider the case of $q=1$. Since the first order moment $\langle M \rangle = 0.5$, we fine that $\tau(1)=0$, and consequently $f(\alpha(1))=\alpha(1)$. Therefore, the $f(\alpha)$ curve is tangent to the diagonal $f(\alpha)=\alpha$ of the first quadrant. This is a universal nature of the $f(\alpha)$ curve extracted via the multiplier method. The probability density of multiplier satisfies Eq. (6) for the base 2. Hence, we have $\langle M \rangle = \langle 1-M \rangle$. Thus $\langle M \rangle = (\langle M \rangle + \langle 1-M \rangle)/2 = 0.5$ and $\tau(1)=0$. More generally, $\langle M \rangle = 1/b$ for any fixed base $b$, which results in the property mentioned above.

For $q=0$, we have $\tau(0)=-1$ and $f(\alpha(0))=D_0=1$. This is an inevitable ending of the construction that there distributes measure everywhere in the geometric support $(0,1)$.

### C. The asymptotic behavior near $q_{\text{bottom}}$ and $q \to +\infty$

In the tail of the right part of the multifractal spectrum, the decay of the experimental curve is much faster than that from the model. This is also universal when comparing multifractal spectra arising from continuous and discrete multiplier probability distribution. When one deal with the experimental data using the multiplier method, the rules of construction one can choose is finite. Hence, $\alpha_{\min} = -\log(\max\{M_i\})/\log b = 1.48 \times 10^{-4}$ and $\alpha_{\max} = -\log(\min\{M_i\})/\log b = 13.25$. However, this is far from being the case when $\Pr(M)$ exists and is continuous. We find that $\alpha \in (0,+\infty)$.

From Eq. (8), $\langle M^q \rangle \sim \dfrac{1.1875}{q+1}$ when $q \to -1^+$ or $q \to +\infty$. Thus we obtain the asymptotic behaviors of the scaling properties near $q_{\text{bottom}}$ and of $q \to +\infty$ as follows.

$$\tau(q) \sim \log_2(q+1). \tag{12}$$

$$D_q \sim -\log_2(q+1)/2. \tag{13}$$

$$\alpha(q) \sim \frac{1}{(q+1)\log 2}. \tag{14}$$

$$f(\alpha) \sim \log_2 \alpha - \alpha. \tag{15}$$

Obviously, $\alpha(q)$ behaves like a hyperbolic function when $q \to +\infty$ or $q \to -1$. Meanwhile, the $f(\alpha)$ curve behaves like a logarithmic function near $q_{\text{bottom}}$ and like a straight line with slope $-1$ when $q \to +\infty$. On the other hand, $f(\alpha)$ has maximum and minimum when utilizing discrete multiplier method, which relates to the refinement $l$ of the construction and $\max\{M_i\}$ or $\min\{M_i\}$.

As proved by Mandelbrot and Evertsz [26], a consequence of $\alpha_{\max} = \infty$ is that $q_{\text{bottom}} = 0$, meaning that for $q<0$ the summation of $q$-order moments fails to scale like $\varepsilon^{\tau(q)}$ and the function $\tau(q)$ fails to be defined. Moreover, a consequence of $\alpha_{\min} = 0$ is that $q_{\text{top}} = 1$. It seems that the resultant definition domain of $q$ in our random multifractal model is in contradiction to the consequences mentioned just now. In fact, a latent condition is used in the proof given in Ref. [26]. The presupposition is that $\alpha(0)=\infty$ for the first consequence and $\alpha(1)=0$ for the second consequence. The proofs can be extended to more general cases. If there exist $q_{\text{bottom}}$ and $q_{\text{top}}$ that satisfy $\alpha(q_{\text{bottom}}) = \infty$ and $\alpha(q_{\text{top}}) = 0$, one can show that $q_{\text{bottom}} < q < q_{\text{top}}$, which leads directly to the result in the present case.

### D. Comparison with lognormal distribution

Deviation of the triangular distribution model from the uniform distribution model is argued in the previous subsection. It is also necessary to say some words about the lognormal distribution, since it is natural for readers to assume that the multipliers $M_i$ fluctuate according to lognormal distribution as in the turbulence model [27,28]. The lognormality hypothesis comes from the application of central-limit theorem to argue that $\log M_i$ should have Gaussian distributions. However, central-limit theorem can't be applied to rare events, which are the ones that contribute most to high-order moments [12]. From the lognormality hypothesis, one may find that the multifractal function $f(\alpha)$ is parabolic in the central "bell" near their maximum and symmetric to $\alpha = \alpha(0)$ [15,29]. We may find that, the multifractal function with lognormal distribution fit these in Fig.6 near $q = 0$ remarkably, but the central-limit theorem does not pretend to say anything when concerning their form away from the maximum. Moreover, we are easy to see these points from Fig.7.

### IV. CONCLUSIONS

In this paper, the multifractal nature of drop breakup in air-blast nozzle atomization process has been studied. We applied the multiplier method to extract the negative and the positive parts of the $f(\alpha)$ curve with the data of drop size distribution measured using Dual PDA, which is base on Taylor's frozen flow hypothesis. On the other hand, we proposed a random multifractal model with the multiplier triangularly distributed to characterize the breakup of drops. The agreement of the left part of the multifractal spectra between the experimental result and the model is remarkable. The cause of the distinction of the right part of the $f(\alpha)$ curve was argued, which is due to the essence of the multiplier method. Comparisons with uniform and lognormal distribution are performed respectively.

The fact that negative dimensions arise in the current system, namely that of $f(\alpha) < 0$, means that the spatial distribution of the drops yielded by the high-speed jet fluctuates from sample to sample. On other words, the spatial concentration distribution of the disperse phase in the spray zone fluctuates randomly momentarily and shows intrinsic randomness. The randomness comes also from the experimental procedure, since the measurement of drop size distribution may be considered as randomly oriented one-dimensional cuts through a three-dimensional spray zone, even if the latter comes from a strictly deterministic process.

### ACKNOWLEDGEMENTS

This research was supported by the National Development Programming of Key and Fundamental Researches of China (No. G1999022103).